\documentclass[11pt]{article}

\usepackage{graphicx}
\usepackage{a4wide} 
\usepackage{latexsym}                    
\usepackage{amsfonts}                    
\usepackage{amssymb}                     
\usepackage{amsmath}                    
\usepackage{hyperref}
\usepackage{booktabs}

\graphicspath{{./figures/}}

\def\R{{\rm I\!R}}
\def\sign{sign}

\usepackage{authblk}
\title{Automated detection and segmentation of non-mass enhancing breast tumors with dynamic contrast-enhanced magnetic resonance imaging}

\author[1]{Ignacio Alvarez Illan}
\author[1]{Javier Ramirez}
\author[1]{J.M. Gorriz}
\author[2]{Maria Adele Marino}
\author[2]{Daly Avendano}
\author[3]{Thomas Helbich}
\author[3]{Pascal Baltzer}

\author[2,3]{Katja Pinker}
\author[4]{Anke Meyer-Baese}

\affil[1]{Signal Theory and Communications Department, Universidad de Granada, Spain}
\affil[2]{Department of Radiology, Memorial Sloan-Kettering Cancer Center, New York, USA}
\affil[3]{Department of Biomedical Imaging and Image-guided Therapy, Division of Molecular and Gender Imaging, Medical University Vienna/AKH Wien, Austria}
\affil[4]{Scientific Computer Department, Florida State University, Tallahassee, FL 32306 USA}

\begin{document}



\maketitle

\begin{abstract}
Non-mass enhancing lesions (NME) constitute a diagnostic challenge in dynamic contrast enhanced magnetic resonance imaging (DCE-MRI) of the breast. Computer Aided Diagnosis (CAD) systems provide physicians with advanced tools for analysis, assessment and evaluation that have a significant impact on the diagnostic  performance.  Here, we propose a new approach to address the challenge of NME detection and segmentation, taking advantage of independent component analysis (ICA) to extract  data-driven dynamic lesion characterizations. A set of independent sources was obtained from DCE-MRI dataset of breast patients, and the dynamic behavior of the different tissues was described by multiple dynamic curves, together with a set of eigenimages describing the scores for each voxel. A new test image is projected onto the independent source space using the unmixing matrix, and each voxel is classified by a support vector machine (SVM) that has already been trained with manually delineated data. A solution to the high false positive rate problem is proposed by controlling the SVM hyperplane location, outperforming previously published approaches. 

\end{abstract}

\section{Introduction}
{A}{ccurate} methods for early diagnosis of breast cancer are pivotal and contribute to an improved prognosis and survival outcomes in breast cancer patients. There is a consensus that dynamic contrast-enhanced magnetic resonance imaging (DCE-MRI) is the most sensitive test for breast cancer detection and the backbone of any MRI protocol, enabling simultaneous assessment of tumor morphology and enhancement kinetics that evaluate neoangiogenesis as tumor specific feature. DCE-MRI has an excellent sensitivity and good specificity for lesions presenting as mass enhancement \cite{jansen_diverse_2011}. However, non-mass-enhancing lesions (NME) exhibit a heterogeneous appearance with high variations in kinetic characteristics and  morphological patterns on DCE-MRI \cite{sakamoto_categorization_2008}. Consequently, DCE-MRI has reported lower specificity and sensitivity of 35\% and 73\% for NME, much lower than that for mass-enhancing lesions. A set of computer aided diagnosis (CAD) systems for breast cancer diagnosis on DCE-MRI has been developed with satisfactory performance results. However, in breast tumors presenting as NME the performance with low specificity is still suboptimal. 

 For a CAD system to be used in breast DCE-MRI two features are important to evaluate: i) the ability of the CAD to correctly differentiate between malignant and benign lesions; and ii) the ability of the CAD system to correctly locate malignant lesions within the 3D spatial volume. To evaluate the first feature, the diagnostic accuracy, specificity and sensitivity are usually reported. To evaluate the second feature, commonly the Dice similarity coefficient (DCS) is calculated between the CAD segmentation and some other ground truth segmentation. In most cases, a manual segmentation of the lesions is performed by experienced radiologists as ground truth. However, it has to be noted that even with expert's interpretation visual readings are prone to subjective errors\cite{pages_undiagnosed_2012} and specificity of DCE-MRI is limited particularly in small and non-mass enhancing lesions, resulting in unnecessary breast biopsies \cite{heywang-kobrunner_contrast-enhanced_1997}. In addition, CAD systems for breast cancer diagnosis have a reported high false positive rate, and consequently low specificity. However, this does not necessarily mean that CAD systems misclassify benign lesions as malignant. Therefore, it is not clear whether CAD systems can be optimized to improve lesion segmentation independently from lesion classification, or if irregardless they will inherently  suffer from the same limitations such as the low specificity reported in visual readings of DCE-MRI.

 In this work we examined the relationship between the false positive rate of CAD systems for breast cancer diagnosis and lesion segmentation on DCE-MRI. To achieve our aim, we obtained rich characterization of data through advanced processing techniques, combined with machine learning paradigms intended for big data analysis and used the resulting information to build a CAD system. We did not introduce any \textit{a priori} knowledge about the disease  in the work flow in order that all information may be completely data driven, which thereby also enabled us to identify new features not currently in the Breast Imaging Reporting and Data System (BI-RADS) classification criteria that could potentially improve segmentation of visual readings. Both morphological and kinetic descriptors are considered in BI-RADs lexicons. However, in NME lesions, morphological descriptors are hard to define and therefore, kinetic behavior can be an important source of information. Therefore, using only dynamic information of the tissue, we performed a supervised method to detect and segment non-mass enhanced lesions on the breast.  
 
Lesion segmentation has been successfully achieved using unsupervised clustering methods,  \cite{mcclymont_fully_2014}, fuzzy c-means (FCM) \cite{chen_fuzzy_2006} or improvements over FCM \cite{jayender_automatic_2014}. In unsupervised clustering, sophisticated pre-processing must be implemented to control the false positive rate, with fine tuning of parameters and/or heuristic steps. On the other hand, it has been demonstrated that processing of dynamic signals provides relevant information for classification of tissues, such as principal component analysis (PCA)-based decompositions  closely related to the 3TP method \cite{eyal_principal_2009}.

Thus, we undertook a combination of supervised segmentation and signal processing to successfully segment NME lesions with control of the false positive rate. Independent component analysis (ICA) was used to extract a set of independent curves that described the possible dynamic behavior of different breast tissues. ICA has been shown to provide richer descriptions of underlying patterns than PCA \cite{illan_18f-fdg_2011,illan_projecting_2010}, and therefore was used for supervised classification in our work. We also incorporated machine learning, whereby we trained a classifier using the information encoded in a whole dataset of subjects, including the dynamic behavior of benign and malignant tissues. Considering features at the voxel level, the system 'learned' to characterize malignant tissues with a support vector machine (SVM). A procedure was implemented to fix the SVM hyperplane location, reducing and controlling the false positive rate. Projecting new unseen data using the unmixing matrix allowed us to obtain the features for estimating the generalization capabilities in a cross validation scheme, and compare it with visual readings of the images reported in the literature and other CAD system approaches. 

The methods proposed within this work demonstrate that NME can be detected with kinetic information by using multiple enhancement curves, providing a promising approach for improving breast cancer diagnosis. Accurate diagnostic methods as the one we hereby present may have an impact not only in accurate diagnosis, but also in reducing unnecessary breast biopsies.

\subsection{Related work}
The use of CAD systems to improve visual readings of DCE-MRI in breast cancer ranges from purely visual methods, to automatic classification. The present work combines visual comparison aspects with automatic classification techniques, thus adding value to purely visual comparison techniques based on PCA or Self-Organizing Map (SOM), such as in  \cite{eyal_principal_2009,varini_visual_2006}, and complementing pure classification approaches, such as in  \cite{gubern-merida_automated_2015,chang_computerized_2014}. Specifically, the PCA approach of  \cite{eyal_principal_2009} extends the three point technique (3PT) by adding an eigenvector decomposition of the time signals. However, that decomposition does not provide an independent set of sources, but only a set of uncorrelated ones. The time-intensity curve estimation of  \cite{liu_total_nodate} also seeks for hidden kinetics, but applies them to mass lesions. Concerning the automatic classification CADs, most approaches are concentrated on the detection and classification of mass enhancing lesions, by combining kinetic and morphological features \cite{hoffmann_automated_2013,gubern-merida_automated_2015,chang_computerized_2014,wang_computer-aided_2014}, like shape, margins, and internal enhancement distribution \cite{agliozzo_computer-aided_2012}, textural kinetic \cite{agner_textural_2011}, or more recently using deep neural networks  \cite{rasti_breast_2017,antropova_su-d-207b-06:_2016}, among others. The detection and segmentation of lesions are usually performed as a manual or semi-manual task, in which regions of interest (ROIs) are manually defined or obtained from seeds with manual inputs.

For automatic lesion segmentation, keeping an acceptable false positive rate is a common issue in DCE-MRI CAD systems of the breast \cite{levman_metric_2016}. In many cases of these cases, unsupervised methods for lesion segmentation, such as FCM algorithms in  \cite{chen_fuzzy_2006,chang_classification_2012}, are used and then the features extracted from the lesions are used for classification. Complex workflows that include vessel detection, whole breast segmentation, and several preprocessing steps have been proposed to control false positive detection \cite{cui_malignant_2009,hu_image_2015,mcclymont_fully_2014,jayender_automatic_2014}.



\section{Methods}

\begin{figure*}[h]
\includegraphics[width=\textwidth]{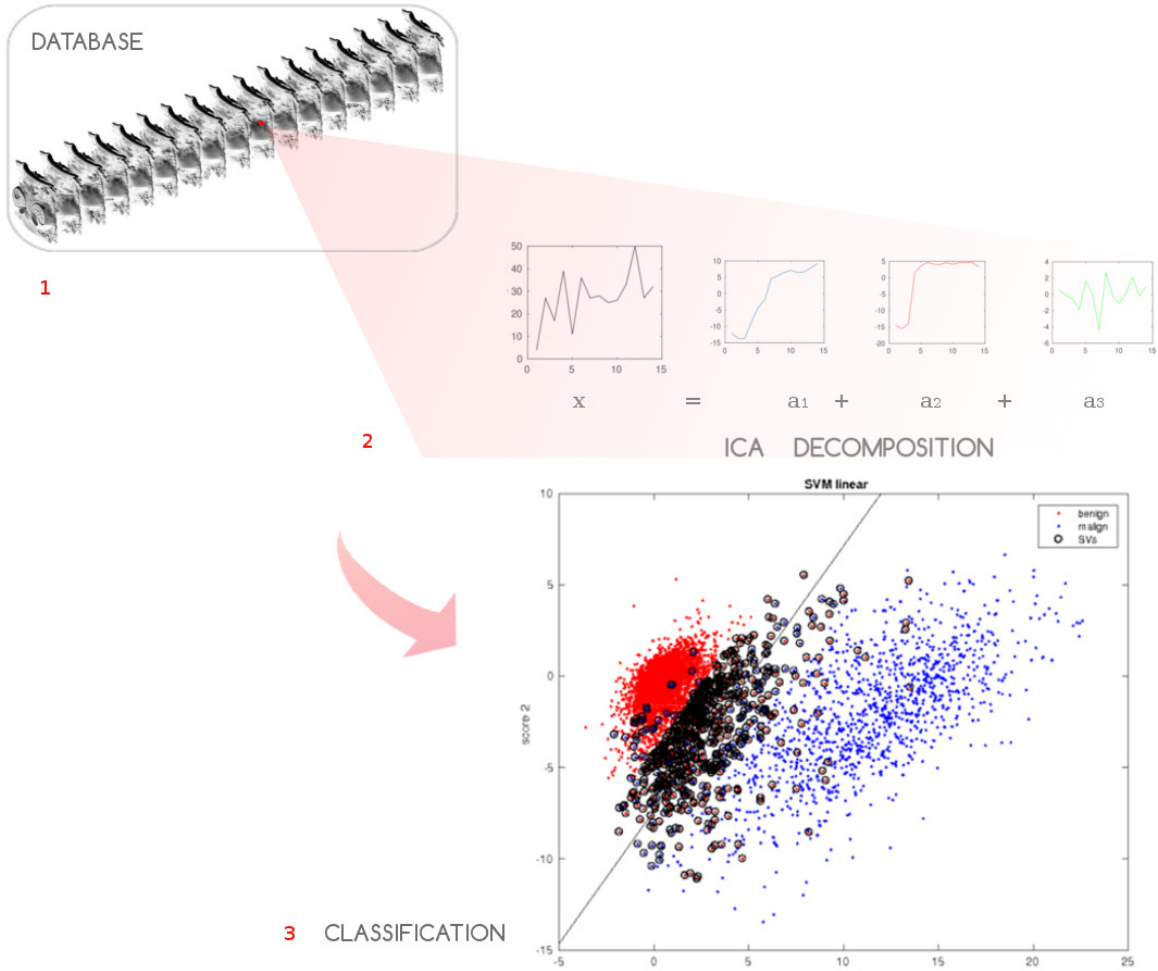}
\caption{1.Time sequence of database images. 2 Decomposition of a sample time signal $\mathbf{x}$ into a linear combination of independent sources by ICA and its corresponding scores $\mathbf{a}_1$, $\mathbf{a}_2$,..., $ \mathbf{a}_s$. 3 Scatter plot of the first scores and the SVM hyperplane classifier.}\label{fig:decomp}
\label{fig:roc}
\end{figure*}

Each voxel of the DCE-MRI image has a time signal representing the enhancement kinetics of the different contributing breast tissues. A set of DCE-MRI time signals can be analyzed in terms of the blind source separation problem, which proposes that the different dynamic behavior can be expressed as a linear combination of a reduced set of sources, making very little assumptions on the nature of that combination. Those sources and their scores can be used as features for classification, as depicted in figure \ref{fig:decomp} 
\subsection{ICA-based enhancement curve analysis}

ICA offers a solution to the blind source separation problem estimating a set of sources that maximizes the statistical independence between them, measured in terms of a cost function. In the literature, several functions have been used to measure statistical independence between signals \cite{Comon94}. Here, we used the fastICA algorithm \cite{hyvarinen_fast_1997} with mutual information as measure function. Contrary to other eigenimage decompositions based on spatial-ICA, like in face recognition \cite{Barlett2002} and brain imaging \cite{illan_18f-fdg_2011,khedher_independent_2016}, the independent sources are obtained here in the temporal domain or, in other words, we work on a voxel level. 

Thus, each voxel defines a temporal curve $\textbf{x}(t_j)$ with $t_1,..,t_N$ temporal points. A set of voxels $\{\mathbf{x}_i\}, i=1,...,M$ forms an image, and defines the $N \times M$ matrix $\mathbf{X}$ of observed signals. The ICA task is to find the mixing matrix $\mathbf{A}$ and the set of \textit{sources} $\mathbf{S}$:
\begin{equation}\label{eq:ICA}
\mathbf{X}=\mathbf{A}\mathbf{S}
\end{equation}
The mixing matrix $\mathbf{A}$ is a $N \times N$ matrix that linearly combines the independent 'images'. Contrary to other related methods, such as PCA, ICA does not provide a natural way to sort the $N$ independent components. However, it is a relevant question whether or not a reduced set $p<N$ of components contain noisy and discardable information. The mean squared error (MSE) between the enhancement time signals and the reconstructed signals using the $k$ source $\textbf{s}_k$, is calculated as:
\begin{equation}\label{eq:MSE}
MSE(k)= \frac{1}{N_t \cdot N_r}\sum_{i,j} (\textbf{x}_i(t_j)-a_{jk} \cdot s_{ki} )^2
\end{equation}
and used as a parameter to measure the noise content of each $\textbf{s}_k$ source, with $k=1,...,N$.

When working at voxel level, equation \ref{eq:ICA} can also be understood as a linear decomposition of each vector $\textbf{x}(t_j)$ into a set of temporal sources whose coefficients belong to the independent sources. Therefore, 
 each voxel location $\{\mathbf{x}_i\}$ has $N$ coefficients $s_{ji}$ to $j=1,...,N$, whose values are maximally independent and measure the importance of each temporal source to recover that voxel dynamics, by linearly combining them (see figure \ref{fig:decomp}). In the rest of the paper, we will refer to these coefficients as the \textit{scores}.

 It is important to stress that working on a voxel level will allow data from different patients to be included in the matrix set $\mathbf{X}$. Therefore, the  obtained set of sources $\mathbf{S}$ does not have to be restricted to represent the particular dynamic enhancement present in a single subject, but can be used to model all the possible curves that independently characterize each BI-RADs category.  

 For new unseen data $\tilde{\textbf{x}}$ at the voxel level, the scores are extracted from $\tilde{\textbf{x}}$ by projecting it onto the subspace $E$ spanned by the signals form the matrix $\mathbf{A}$. Specifically, let $\{\textbf{a}_1, ..., \textbf{a}_p\}$ be the basis set of temporal curves spanning the subspace $E$, then $\mathbf{A}$ denote the $N$-by-$p$ matrix of which columns are $\textbf{a}_1, ..., \textbf{a}_p$. Let $p\leq N$, as some of the signals may have been removed due to its noise nature. Since this basis need not be orthogonal, a well known result of linear algebra stated that the projection is given by:

\begin{equation}
\mathbf{P}_A = \mathbf{A} (\mathbf{A}^T \mathbf{A})^{-1}\mathbf{A}^\mathrm{T}
\end{equation}

\noindent so that the application of that operator on a voxel signal $ \tilde{\textbf{x}}(t)$:
\begin{equation}\label{eq:proj}
\mathbf{s}=\mathbf{P}_A\tilde{\textbf{x}}
\end{equation}

\noindent projects it to the subspace $E$, obtaining its $p$ scores $\mathbf{s}$ on that subspace. 

The independent component scores  $\mathbf{s}_k$ of the dataset is used as feature vectors inputs of a SVM to learn the different enhancement patterns associated with malignant and benign tissues. 

\subsection{False positive rate control by SVM hyperplane translation}

SVM is a machine learning algorithm that separates a given set of binary labeled training data with a hyperplane that is maximally distant from the two classes (known as the maximal margin hyper-plane). The objective is to build a function $f:\R^{p} \longrightarrow \{1,0\} $ using training data, consisting of \emph{p}-dimensional patterns $\textbf{\texttt{x}}_i$ and class labels $y_i$:

\begin{equation}\label{training_data}
(\textbf{x}_1,y_1), (\textbf{x}_2,y_2), ..., (\textbf{x}_M,y_M)  \in \left(\R^{p}  \times \{1,0\}\right),
\end{equation}
\noindent so that \emph{f} will correctly classify new examples $(\tilde{\textbf{x}},y)$. The problem of finding the maximal margin hyperplane is usually solved by quadratic programming algorithms that try to minimize a margin cost function $J$:
\begin{equation} \label{cost_function_no_separable}
J(\textbf{\emph{w}},w_0,\xi)= \frac{1}{2}||\textbf{\emph{w}}||^2+C\sum_{i=1}^l\xi_i,
\end{equation}
subject to the inequatity constraints:
\begin{equation} \label{constr_no_separable}
y_i[\textbf{\emph{w}}^Tx_i+w_0]\geq1-\xi_i, \quad \xi_i \geq 0 \quad i= 1, 2, ..., l.
\end{equation}

\noindent where the slack variables $\xi_i$  incorporate to the optimization those feature vectors that are not separable (details can be found in  \cite{Vapnik1998}). The solution to that problem can be expressed by a linear combination of a subset of vectors, called support vectors: 
\begin{equation}\label{eq:non_linear_svm}
d(\textbf{x})= \sum_{i=1}^{N_S}\alpha_iy_i K(\textbf{s}_i,\textbf{x})+w_0\,
\end{equation}
where $K(.,.)$ is the kernel function, $\alpha_i$ is a weight constant derived form the SVM process and $\textbf{s}_i$ are the $N_S$ support vectors  \cite{Vapnik1998}. Taking the sign of the function leads to the binary classification solution.

Here, we propose a SVM hyperplane translation in terms of the slack variables $\xi_i$, to control the number of false positives. We add a new term $g(\textbf{s}_i,\xi_i)$ to the hyperplane defining function $d(\textbf{x})$, so that the classification solution is now defined by:

\begin{equation}\label{eq:dist}
f(\textbf{x})= \sign\{d(\textbf{x})+g(\textbf{s}_i,\xi_i)\}
\end{equation}

\noindent where the function $g$ takes the  two-class average distance to the hyperplane of those support vectors with $\xi_i>1$, measured by the kernel metric $K$.
Common kernels that are used by SVM practitioners for the nonlinear feature mapping are:

\begin{itemize}
    \item Polynomial
\begin{equation}\label{pol_kernel}
K(\textbf{\emph{x}},\textbf{\emph{y}})= [ \gamma (\textbf{\emph{x}} \cdot \textbf{\emph{y}}) + c]^d.
\end{equation}
    \item Radial basis function (RBF)
\begin{equation}\label{rbf_kernel}
K(\textbf{\emph{x}},\textbf{\emph{y}})= \exp(-\gamma||\textbf{\emph{x}}-\textbf{\emph{y}}||^2).
\end{equation}
\end{itemize}
as well as the linear kernel, in which $K(.,.)$ is simply a scalar product, and therefore $g$ in equation \ref{eq:dist} would average the euclidean distance in that particular case.

\subsection{Dataset}

The dataset used for analysis consisted of sixteen patients that presented with NME breast tumors at DCE-MRI. This patient cohort is a subset from a larger cohort undergoing multiparametric MRI using inclusion criteria described in detail in  \cite{pinker_combined_2009}. All patients underwent MRI of the breast using a 3T MRI scanner(Tim Trio, Siemens, Erlangen, Germany) with a dedicated, bilateral, 4-channel breast coil in vivo (Orlando, FL) and the imaging protocol comprised both high spatial and temporal resolution. Three high spatial resolution images were taken, pre-contrast, peak, and postcontrast as coronal T1-weighted (3D) FLASH sequence, with water excitation and fat suppression, with the following sequence parameters: TR/TE 877/3.82 milliseconds; FOVr 320 mm, SI 1 mm isotropic, 96 slices, flip angle 9$^{\circ}$ , matrix 320 / 134, 1 average, acquisition time 2 minutes. High temporal resolution, contrast-enhanced, coronal T1-weighted (VIBE) sequence was obtained with the following sequence parameters: TR/TE 3.61/1.4 milliseconds, FOVr 320 mm, SI 1.7 mm isotropic, 72 slices, flip angle 6$^{\circ}$ , matrix 192 / 192, 1 average, 13.2 seconds acquisition time per volume leading to 3.45 minutes for 17 measurements. A second set of high spatial resolution T1-weighted imaging (repeated 3D-FLASH) was acquired after these 17 low spatial VIBE resolution images, as the peak enhancement of the lesion could be expected at the end of this time span( \cite{pinker_combined_2009} and references therein). Finally, high temporal resolution (repeated VIBE with 25 measurements, leading to an acquisition time of 5 minutes 35 seconds and repeated 3D-FLASH for dynamic assessment of lesion wash-out) was performed, and then high spatial resolution T1-weighted images were recorded. The contrast agent used was Gd-DOTA (generic name: Gadoterate meglumine; Dotarem, Guerbet, France), injected intravenously as a bolus (0.1 mmol per kilogram body weight) and administered with a power injector (Spectris Solaris EP, Medrad, Pittsburgh, PA) at 4 mL/s followed by a 20 mL saline flush. The contrast agent was injected 75 seconds after starting the first coronal
T1-weighted VIBE.

NME breast tumors were visually assessed by three expert radiologists following the American College of Radiology BI-RADS atlas \cite{Bi-rads}, and delineated using the Osirix software on the 3T high spatial resolution volumes. All NME were classified as BI-RADS 4: suspicious or BI-RADS 5: highly suspicious of malignancy. Histopathology was used as the standard of reference. There were eleven invasive ductal carcinoma (IDC), three ductal carcinoma in-situ (DCIS) and two invasive lobular carcinoma (ILC). 

\subsection{Preprocessing}

All dynamic sequences were registered to the pre-contrast volume. This pre-processing step was required to remove any spatial misalignments on the sequence caused by involuntary movements of the patient. The algorithm employed to perform this task was the SPM12 \cite{spmbook} registration algorithm, which performs affine and non-affine transformations on the data by minimizing a similarity measure cost function, selected to be the mutual information metric. Afterwards, a 3D Gaussian filter of size 2FWHM was used to smooth the images. 

\begin{figure}[htb]

\begin{minipage}{0.47\linewidth}
  \centering
  \centerline{\includegraphics[width=\linewidth]{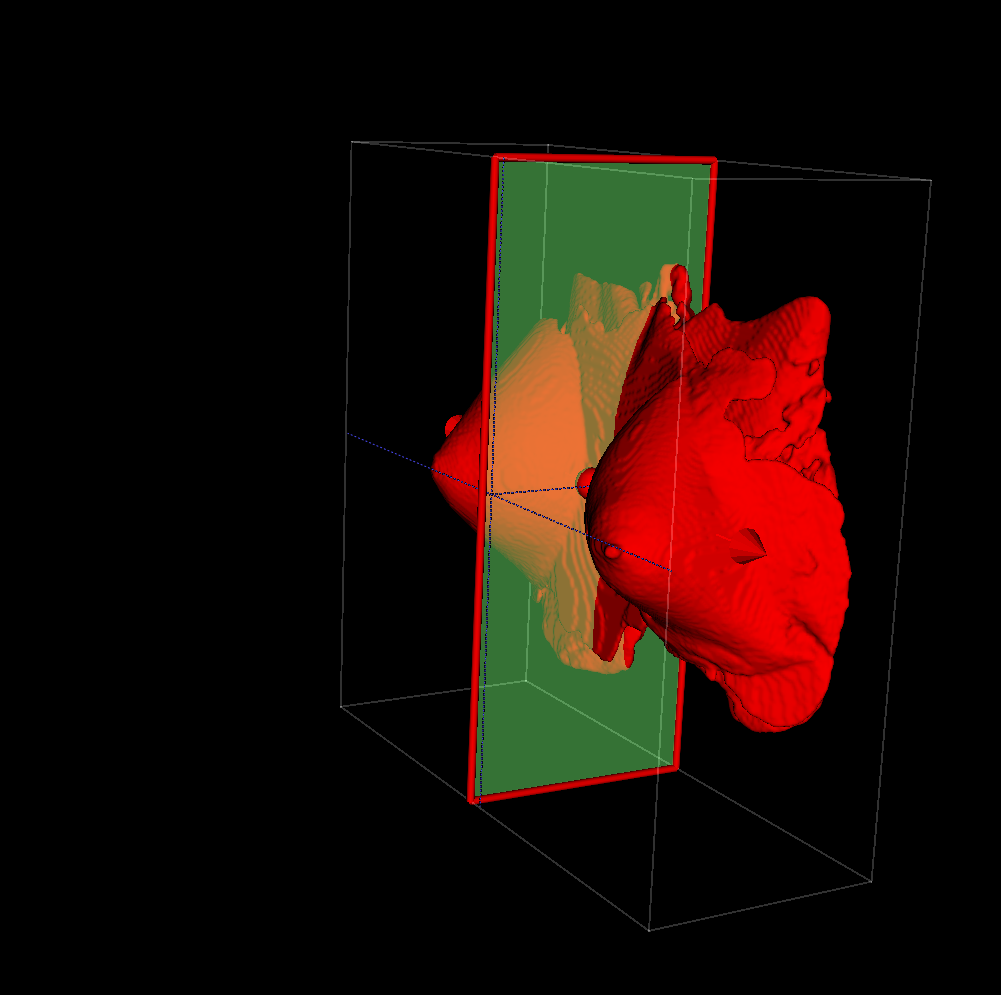}}
  \centerline{(1) Middle sagittal plane}
\end{minipage}
\hfill
\begin{minipage}{0.47\linewidth}
  \centering
  \centerline{\includegraphics[width=\linewidth]{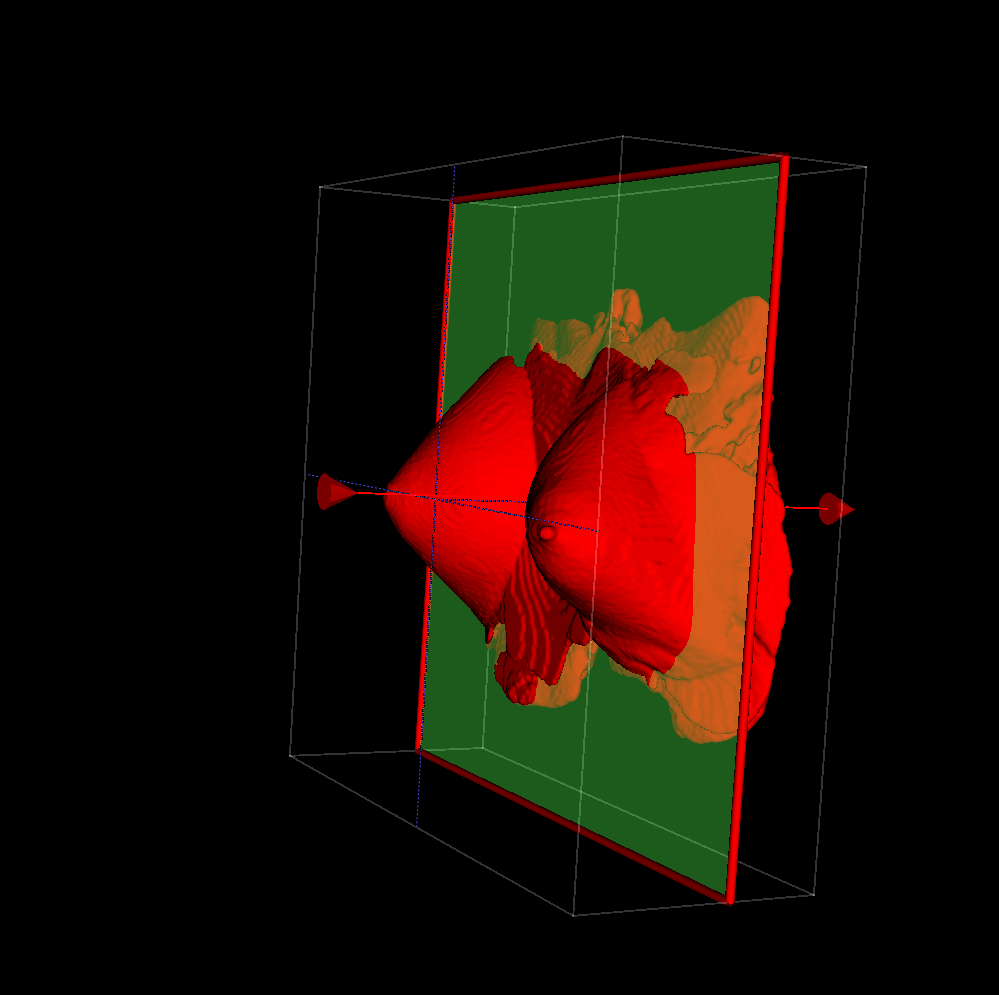}}
  \centerline{(2) Middle chest plane}
\end{minipage}\caption{Exclusion of internal organs by detection of middle chest plane.}\label{fig:midplanes}
\end{figure}

In spite of the existence of automatic and accurate methods for performing whole breast segmentation \cite{dalmis_using_2017,gubern-merida_breast_2013,jiang_fully_2017,wu_automated_2013}, we performed this task straightforwardly finding the middle chest point as in  \cite{chang_computerized_2014}, and discarding the content of the image after this point, reducing the original number of $192 \times 192 \times 72 \approx 2.6 \cdot 10^6 $ voxels contained in each image to $\approx1.6\cdot10^5$, and guaranteeing the exclusion of heart and other organs noisy-signals. Concretely, the middle chest point was obtained by performing the following steps(see figure \ref{fig:midplanes}):
\begin{enumerate}
\item Compute the cross-correlation of the convolution of the image with itself in the sagittal direction. The middle sagittal plane will lie in the symmetry plane of the body, and due to its symmetry it will reach the maximum convoluted cross-correlation.
\item Compute the intensity gradient of the middle chest slice in the coronal direction and find its maximum $m_y$. Remove the internal part of the image that lies in the coronal direction after the middle chest plane $y=m_y$.
\end{enumerate}

The described procedure ensured the removal of voxels that lie inside the thoracic cavity and the chest wall as well as background voxels.

The manual delineations of the lesions were performed by three expert radiologists on 3T high-spatial resolution images using the Osirix software, recorded as a set of axial point coordinates in mm. The Bresenham algorithm \cite{bresenham_algorithm_1965} was used to transform the coordinate points into 3D binary masks, and a decimation was employed to downsample the masks to the size of the low spatial high-temporal resolution images. Thus, the downsampled masks were used to define the class labels of each voxel, 1 if the voxel was in the mask, and 0 otherwise.

 \section{Experiments}
\label{sec:experiments}

The dataset was divided into three subsets: training data, validation data and testing. Training and validation data comprised half of the dataset, while the test set consisted of the other half.  The data were considered at the voxel level.  Therefore, after discarding non-relevant parts of the image, a random selection of $N_a\approx 5 \cdot 10^3$ benign voxel samples from the pool of all non-lesion voxels of the images was performed to balance the training set, resulting on a $2*N_a \times p$ training and validating data matrix. 

The voxel data were used as in input to the fastICA algorithm, obtaining a set of scores for each voxel that served as feature vectors for training and validating a SVM in a cross validation scheme. The validation step is preformed in two stages:

\begin{itemize}
\item First, different parameters were optimized within a 10-fold cross-validation scheme: i) the optimal dimensionality of the data $h$; ii) the optimal kernel (linear, polynomial or RFB). The optimal value for $h$ was obtained by sorting the independent components by their MSE defined in equation \ref{eq:MSE}, and the feature space dimension was changed by sequentially increasing the number of components included on the scores. The optimal kernel was selected by comparing the classification performance, based on the classification error. 
\item Secondly, once the number of components and the kernel function were fixed, the decision boundary location of the SVM was analyzed in an enlarged test dataset of size $\approx 4 \cdot 10^5$, that contained all the discarded voxels in the validation step. 
\end{itemize}


\section{Results}
\begin{figure*}[h]
\centering
\includegraphics[width=\textwidth]{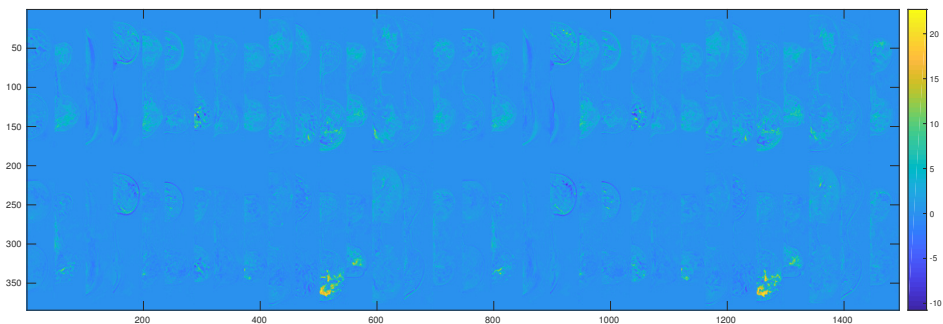}%
\caption{One middle axial slice for the 16 patients from the projected independent component space of the dataset. Intensity represents voxelwise scores of the first and second independent component in the 3D MRI space. The scores of the first IC, on the top, correspond with the IC1 in blue in figure \ref{fig:ics}. The scores of the second IC, on the bottom, correspond with the IC2 in red in figure \ref{fig:ics}. High values on the IC2 (in yellow) can be related to malignancy.}\label{fig:icsmap2}
\end{figure*}

\begin{figure}
\centering
\includegraphics[width=\textwidth]{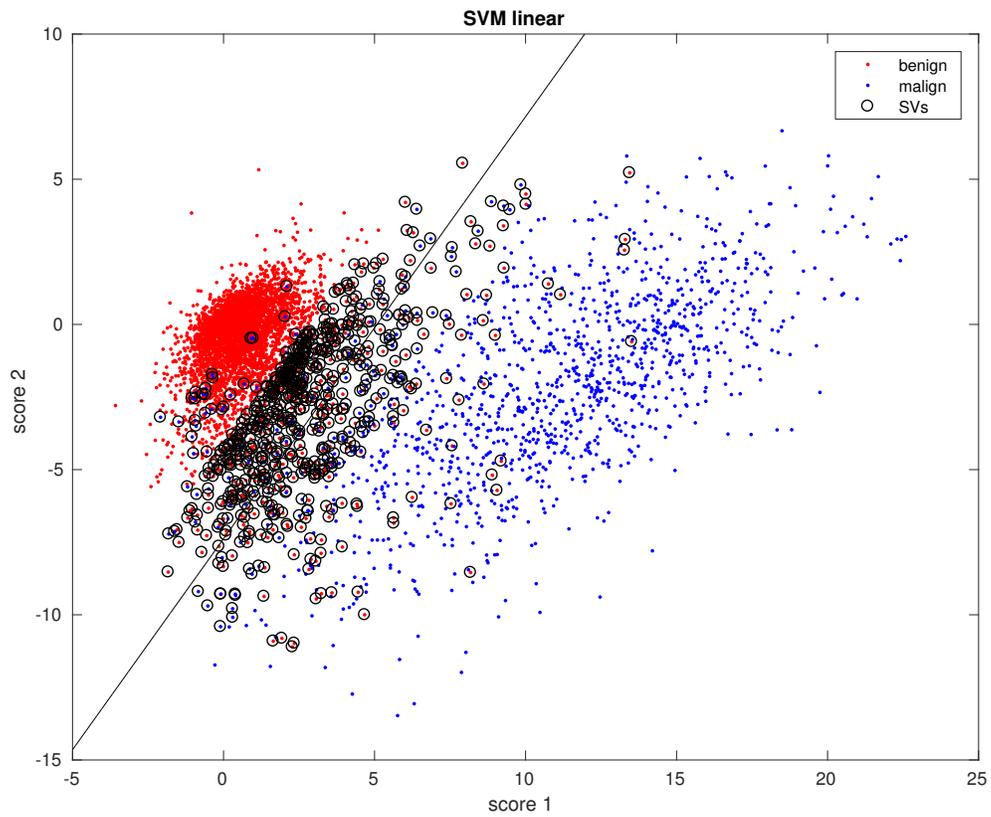}%
\caption{Scatter plot of the scores corresponding to the two first independent components of the training data, together with the linear decision SVM function (in black) and the support vectors (SVs).}
\label{fig:SVM}
\end{figure}

\begin{figure}[h]
\centering
\includegraphics[width=\textwidth]{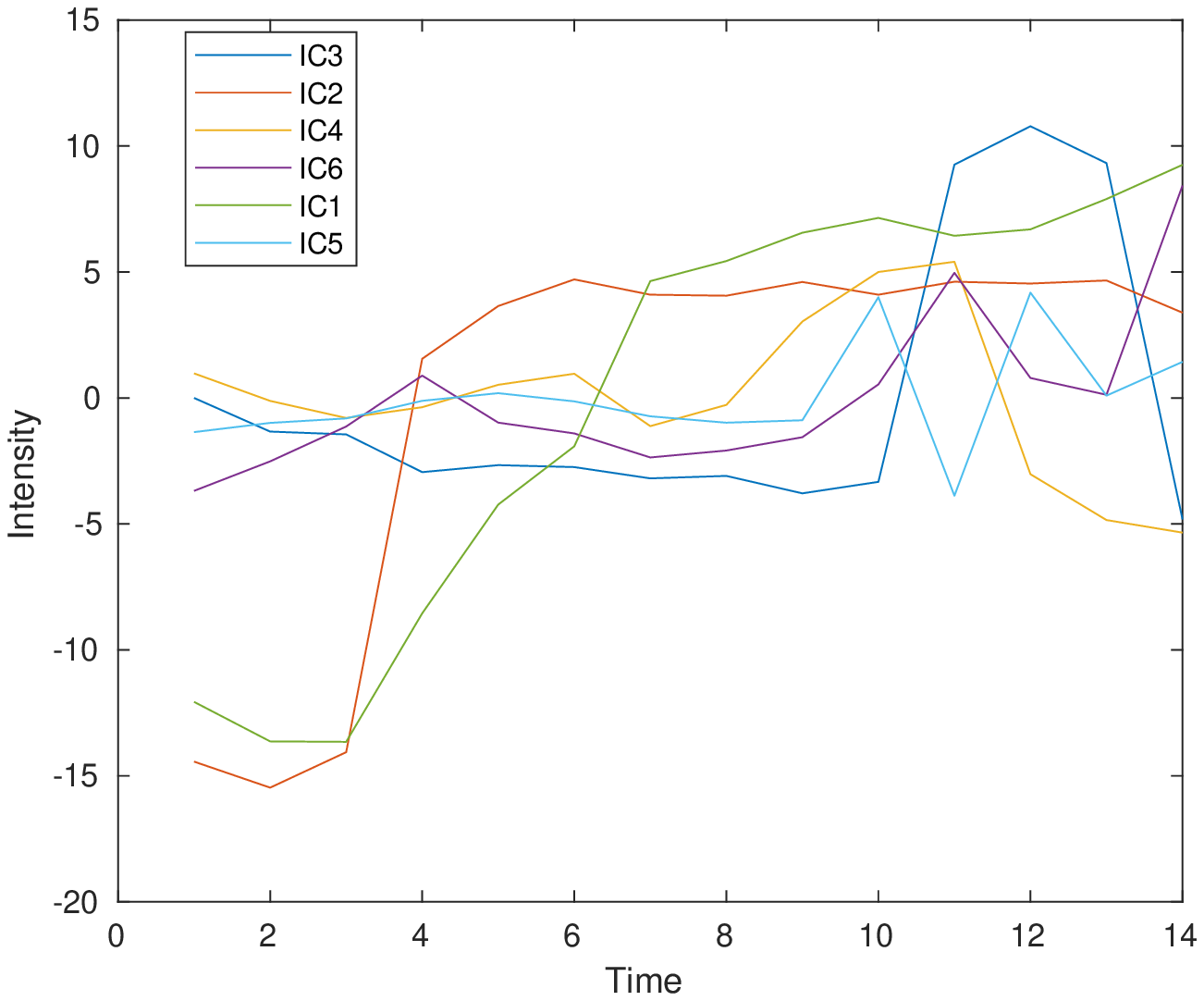}%
\caption{First four independent components sorted by MSE. In red, the IC2 shows typical 'malignant' dynamics, while in blue, IC1 shows a persistent enhancement curve, characteristic of benign tissues. }\label{fig:ics}
\end{figure}

The scores defined in equation (\ref{eq:ICA}) are depicted in two different spaces: the 3D DCE-MRI space corregistered with the original data (figure \ref{fig:icsmap2}), and the $E$ subspace spanned by the two first temporal sources $\mathbf{a}_1$ and $\mathbf{a}_2$ (figure \ref{fig:SVM}), sorted according to the MSE defined criteria. 
The representation in the 3D DCE-MRI space shows that similar score values are grouped together around tissues that have a similar enhancement. On the bottom, voxels belonging to the lesions present a high score value, revealing that the associated independent component encodes the malignant dynamic information. On the top, the distribution of score values does not concentrate in specific regions, but spreads over the breast tissues revealing a relation with normal tissue enhancement dynamics. That information  complements the representation on the $E$ subspace, where a clear separation between tumor tissues represented in blue, and normal tissues in red, can be inferred, although some region of overlapping is present. Also, the 
independent components $\mathbf{s}_1$ and $\mathbf{s}_2$ are shown in figure \ref{fig:ics}, together with other extracted sources. It is interesting to note that, being automatically data-driven extracted, these independent components take the form of enhancement curves: curve IC1 is a normal enhancement, while curve IC2 has a 'typical' malignant behavior, according to model based descriptions \cite{jansen_diverse_2011}. The remaining set of independent components cannot be assigned to any particular dynamic, nor tend to form clusters of similar enhancement when depicted in 3D, therefore not possessing an obvious interpretation. However, the common classification onto wash-out, plateau and permanent enhancement of dynamic curves is reduced by ICA to only two clearly identifiable curves. Therefore, the ICA based signal processing analysis reveals that dynamic enhancement curves reaching a plateau do not behave independently in the ICA sense from washout curves, while permanent enhancement curves do. 

\begin{figure}[h]
\centering
\includegraphics[width=\textwidth]{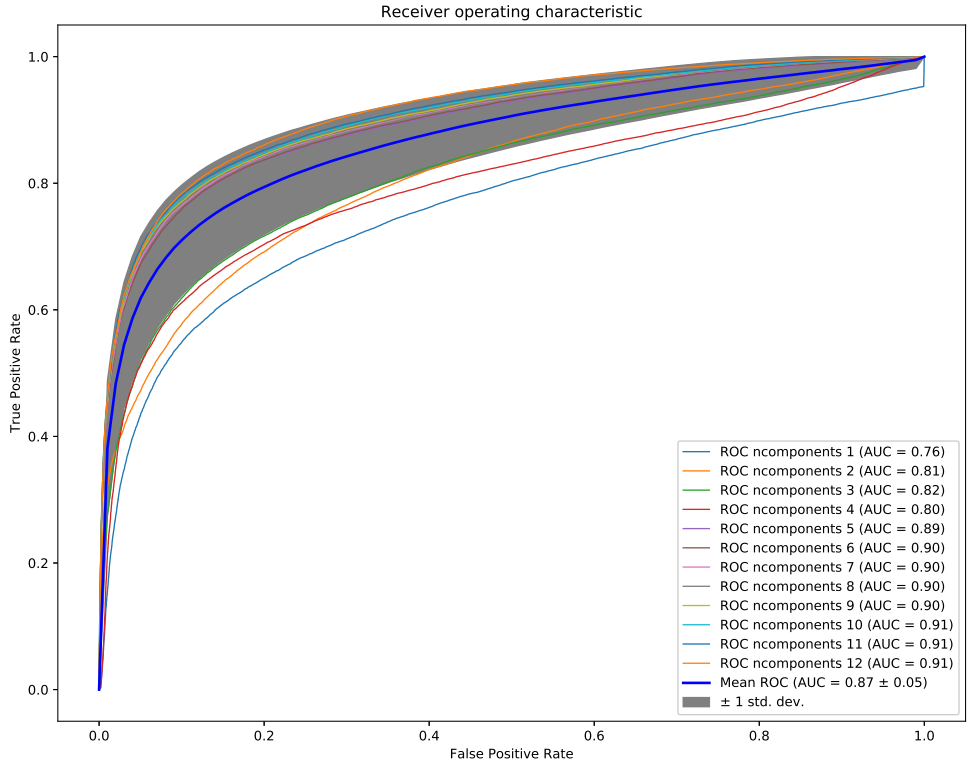}%
\caption{ROC and area under curve (AUC) values on the cross-validation scheme for RBF kernel by varying the number of components on ICA. ICA components are sorted according to MSE.}
\label{fig:err}
\end{figure}

The results of the cross-validation are shown in figures \ref{fig:SVM} and \ref{fig:err}, and also in the left part of table \ref{tab:the_table}. In figure \ref{fig:SVM}, the  $2*N_a \times p$ training data are shown after the SVM is trained, and the obtained support vectors are marked with circles. From figure \ref{fig:err}, the optimal number of component used to reconstruct the signal is above 5, revealing that a simple decomposition of signals into benign and malignant behavior can be enriched with other significant components reaching ROC values over 0.90.  

\begin{figure}[h]
\centering
\includegraphics[width=\textwidth]{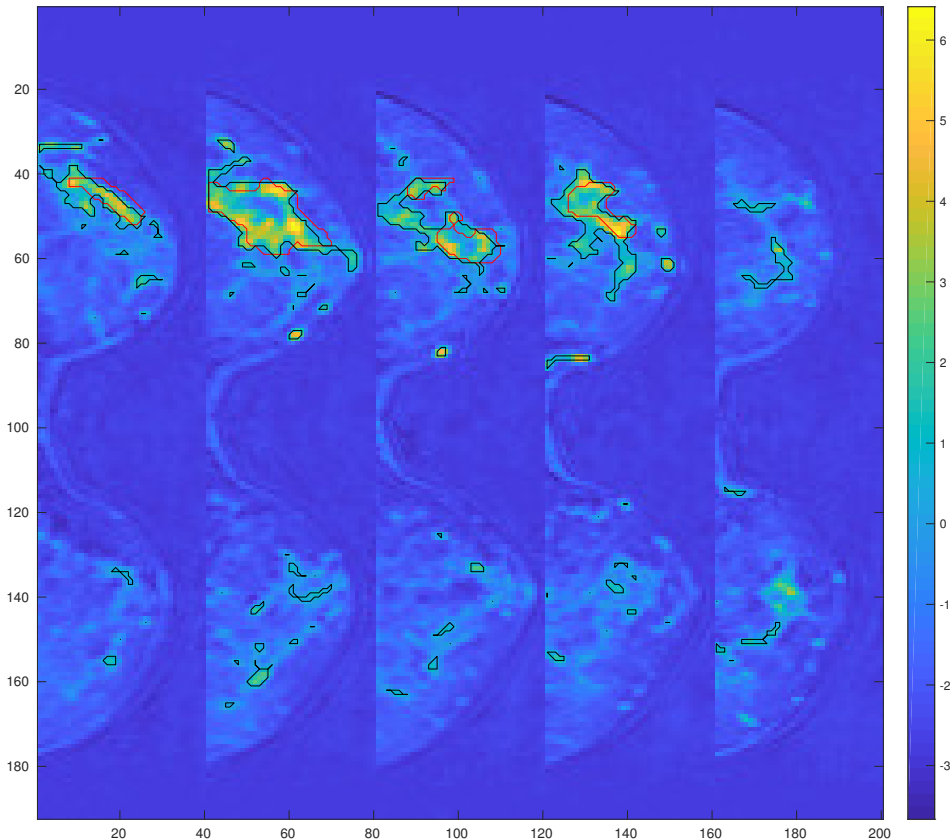}%
\caption{Five representative axial slices of a NME. The values on voxels represent the distance to the hyperplane after classification on a trained SVM. The black contour represents the location of the hyperplane at $d$=0 and the red contour is the manual delineation of the lesion. }
\label{fig:hyp}
\end{figure}

\begin{figure}[h]
\centering
\includegraphics[width=\textwidth]{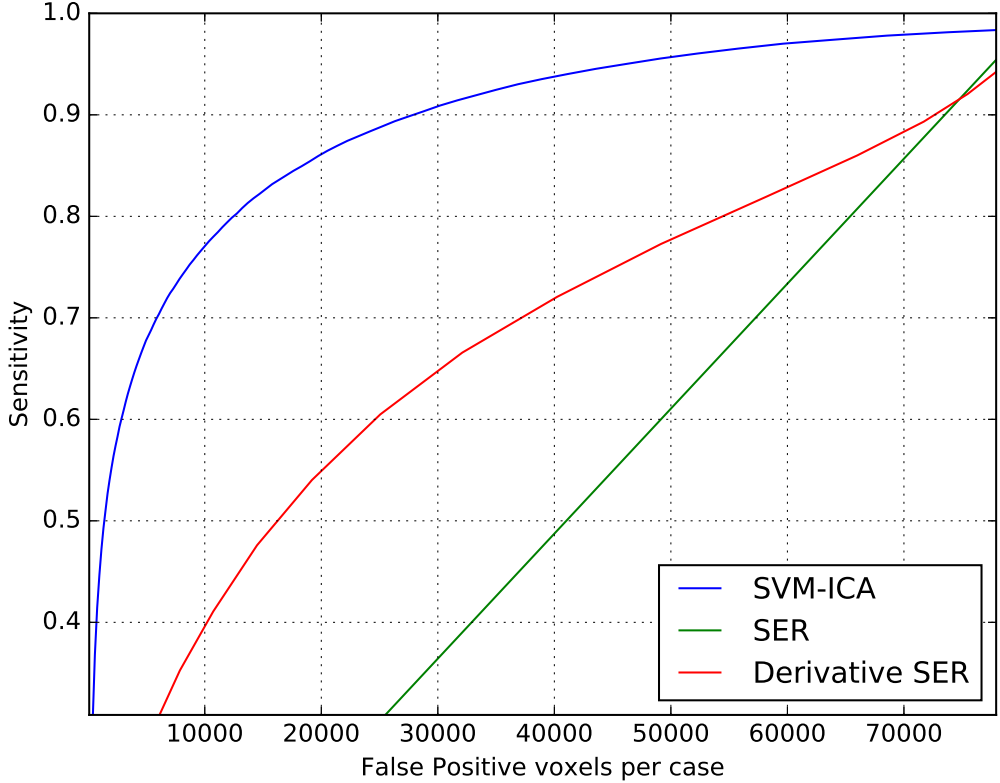}%
\caption{FROC curves for the proposed algorithm (SVM-ICA) in comparison with the references (SER and SER-derivative\cite{levman_metric_2016}).}
\label{fig:dcs}
\end{figure}

\begin{figure}[ht]
\centering
\includegraphics[width=\textwidth]{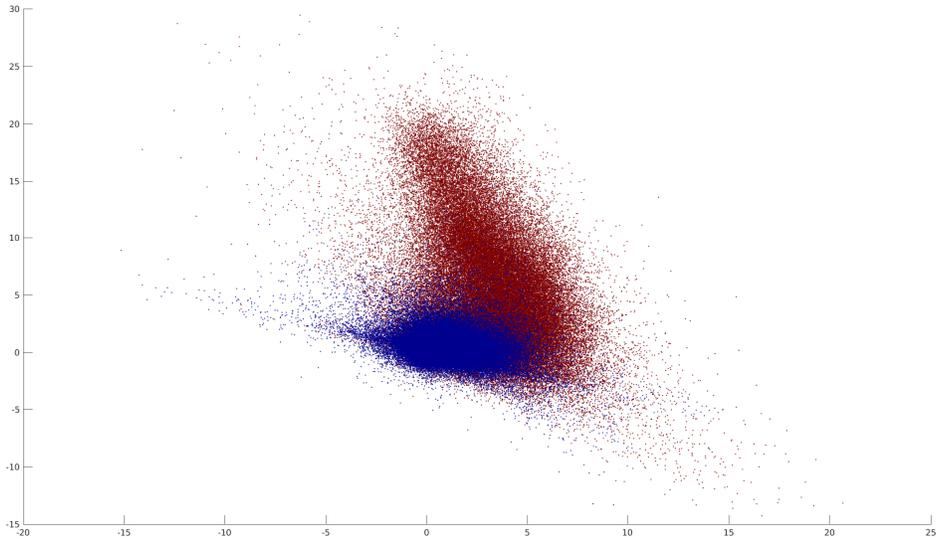}
\caption{Scatter plot of the scores corresponding to the two first independent components of the validation data.}
\label{fig:scatter}
\end{figure}

\begin{table*}[h]
    \centering
    \caption{Performance parameters on training and validating data}
    \label{tab:the_table}\resizebox{\textwidth}{!}{
    \begin{tabular}{ c  c c cc  c c c c }
        \toprule
                  & \multicolumn{4}{c}{Training }& & \multicolumn{3}{c}{Validation DSC}\\ 
\cmidrule{2-5} \cmidrule{7-9}                  
        & Hinge Loss & Accuracy & Specificity & Sensitivity  & & d=0 & max  &       DSC($\mu_d$) [DSC($\mu_d \pm \sigma_d$)] \\ \midrule
                        
PCA + linear SVM& 0.9764 & 0.7263 & 0.6581 & 0.7944& &0.31$\pm$0.01 &0.3382$\pm$0.0005 &0.3310 [0.3039-0.2169]\\
 PCA + RBF SVM      & 0.9529 & 0.7263 & 0.6581 & 0.7944& & 0.31$\pm$0.01&0.3382$\pm$0.0005 &0.3310 [0.3039-0.2169]\\ \midrule
  ICA + linear SVM  & 0.1254 & 0.9501 & 0.9410 & 0.9593& & 0.31$\pm$0.01&\textbf{0.53$\pm$0.01} &  \textbf{0.5295 [0.475-0.484]}\\ 
  ICA + RBF SVM     & 0.1083 & 0.9515 & 0.9573 & 0.9457& & 0.29$\pm$0.01&0.44$\pm$0.04 &0.1085 [0.3711-0.0559]\\ \midrule
  raw + linear SVM  & 2.4429 & 0.8026 & 0.8446 & 0.7605& &  0.15$\pm$0.07&0.30$\pm$0.05 &  0.2325 [0.1373-0.3058]\\ \bottomrule
    \end{tabular}}
\end{table*}

Figure \ref{fig:hyp} shows the NME delineated by the expert radiologist (in red), together with a distance-to-hyperplane map (distance $d=0$ is represented by a black contour). The value of each voxel in the map is defined in equation \ref{eq:non_linear_svm}. It can be seen that hyperplane location (value $d=0$) produces  big regions of false positives. Those regions are mostly concentrated around the delimited lesion, but extended regions can also be found in non-connected regions where benign dynamics are expected. The false positive rate can be controlled by modifying the defining value of the hyperplane location, set to 0 by definition in SVM. Translating the hyperplane towards the positive values produces a more conservative definition of feature vectors belonging to the +1 class.  Therefore, only score values high above the hyperplane would be considered as malignant, while intermediate values not clearly projecting malignant-related score values will not be classified as lesion, decreasing the false positive rate and increasing specificity. However, there must be a compromise between specificity and sensitivity, since increasing the defining value of the decision function also has an impact on the false negative rate. This trade-off requires to be very finely tuned, as there exists a big disequilibrium between the number of +1 class samples and the number of -1 class samples. The several-orders of magnitude bigger number of benign samples produce an imbalance in the validation of the SVM classifier. In figure \ref{fig:scatter}, the influence of the imbalanced classes can be perceived if compared with the scatter plot of the scores considering only the reduced training data of figure \ref{fig:SVM}. Although other solutions exists to the problem of imbalanced dataset in SVM classification, we propose here a very conservative approach, in which the hyperplane defining value is translated into the +1 class region, guaranteeing that only very distant scores from the hyperplane are considered as malignant. The value into which the hyperplane is rotated DSC($\mu_d$) is defined in equation \ref{eq:dist}. Other values could be used to make this transformation, but are prone to be affected by outlier support vectors that uncontrollably increase the false negative rate. By averaging the support vector's distance to the hyperplane with the condition $\xi_i>1$, we are smoothing the effect of possible outlier support vectors, while translating the hyperplane to actual relevant values.  Alternatively, we calculate the decision defining value experimentally, in the second validation on the training data, and test both on the test set: the theoretically derived  value and the experimentally adjusted one. In the special case in which all $\xi_i$ are less than 1, we average the support vector's distance to the hyperplane with the condition $1>\xi_i>0$. 

 To evaluate the  lesion detection performance, the DSC is calculated, defined as:
\begin{equation}
DSC= 2*\frac{A\bigcap M}{A \bigcup M}
\end{equation}
\noindent and measures the amount of overlap between segmentation algorithms (A) and manually-generated (M) segmentations with respect to the size of the segmented region. 

Table \ref{tab:the_table} shows the validation values obtained by default SVM at $d=0$, at empirical maximum, and at proposed value, for 2 component PCA, ICA and raw data using 2 kernels. Raw data is displayed for reference, and corresponds to the use of dynamic curves as feature vectors for SVM, without multicurve extraction. PCA method \cite{eyal_principal_2009} shows higher DSC at $d=0$ than the proposed ICA approach. Hyperplane translation has a lower effect in the PCA case since all support vectors lie in the $\xi_i<1$. In the ICA with a linear kernel case, the false positives are reduced significantly reaching the maximum DSC values, in agreement with the interval of maximum empirical values.

Figure \ref{fig:dcs} reports a free-response receiver operating characteristic (FROC) curve analysis\cite{chakraborty_free-response_1990} at the voxel level. Although in mass lesion FROC analysis is usually reported at lesion level, in NME lesions FROC analysis at lesion level can be misleading, as can be seen from figure \ref{fig:hyp} : increasing the confidence threshold increases the number of false positive lesions due to lesion fragmentation, although false positives at the voxel level decrease. Two reference methods are shown for comparison: The signal enhancement ratio (SER) method, based on the following definition: SER=(SI(t=1st post contrast time point)−SI(t=pre-contrast))/(SI(t=final post contrast time point)−SI(t=pre-contrast)), with a varying threshold; and the Derivative-SER, a modified version of the method that uses the Laplacian of the image to obtain the SER, as defined in the work of Levman et. al.\cite{levman_metric_2016}. The FROC curve for the ICA-SVM method proposed in this paper is obtained on the test set by adding a varying threshold $k$ to the SVM output in eq. \ref{eq:non_linear_svm}, and compute the sign $\sign\{\textbf{d}(x)+k\}$.


\section{Discussion}

The contributions of this work are two-fold: First, visual interpretations of DCE-MRI image can be enriched by using the proposed ICA-based processing of time signals, which produces a data-driven decomposition of dynamic enhancement signals into a multi-curve description, that are statistically independent and disease specific. The idea of producing multiple-curves to characterize lesions has also been explored by Lui et. al.  \cite{liu_total_nodate}, but from the total variation perspective, which is not data-driven but based on assumptions on the data. Other visual methods based on CAD techniques, as PCA in Eyal et. al.  \cite{eyal_principal_2009} or PCA-SOM-LDD in Varini \cite{varini_visual_2006} have been proposed in the literature, to enrich the well-known 3TP method. Thus, visual support is an important characteristic to evaluate in aiding diagnosis of breast cancer by computer systems. It is also important to stress that the ICA extraction must be done only in the training phase of the algorithm. The CAD system will then benefit from an \textit{online} response, once the CAD is conveniently trained. The presented approach outperforms PCA based methods as shown in table \ref{tab:the_table} in terms of automatic segmentation performance, and provides a meaningful visual support for experienced and unexperienced readers. 

The low incidence of NME lesions reduces the available testing data, therefore limiting the validation of the presented method. Moreover, the heterogeneous nature of NME lesions also limits the accuracy in lesion annotation performed by experts when compared to CAD segmentations. Therefore, the reported DSC values when comparing ground truth and CAD results must be understood as a lower bound estimation of the segmentation capabilities of the presented CAD, since a semi-automatic annotation can potentially boost the DSC values. 
 
The second contribution is the \textit{supervised} nature of the detection and segmentation method, which allows control of the false positive rate. Most CAD systems for lesion classification start from a manual or semi-manual ROI deliniation\cite{eyal_principal_2009,chang_computerized_2014,wang_computer-aided_2014}, that limits control of the false positives. The baseline approach to lesion segmentation is the FCM unsupervised method, which in Liang et. al. \cite{liang_lesion_2012} is reported to have a  6\% $\pm$  9\% of overlap with manually defined ROIs, and is commonly used in many CAD systems for breast cancer diagnosis in DCE-MRI. In Jayender et. al. \cite{jayender_automatic_2014}, an enhancing pre-processing step is added to the usual FCM algorithm using linear dynamic system modeling. The overlap of the algorithm output with the radiologists' segmentation and CADstream output, computed in terms of DSC, was 0.77 and 0.72 respectively. In the unsupervised approach of Cui et al \cite{cui_malignant_2009}, a combination of Gaussian mixture modeling and a marker-controlled watershed transform was used to segment the lesions. The overall overlap ratio between the two radiologists' manual segmentations and the proposed algorithm was 64.3\% $\pm$ 10.4\%. The supervised method of Liang et. al. \cite{liang_lesion_2012} shows overlap rates with the ground truth of 51\% $\pm$ 26\% and 48\% $\pm$  25\%. This method required a robust intensity normalization method to make intra-patient comparisons, while the ICA-method presented here characterizes the form of the curve, thus not requiring intensity normalization. Moreover, we report higher or comparable DSC values than those in the literature, even in the more challenging case of NME breast lesions. We also report better control of false positive rate than the method proposed by Levan et. al.\cite{levman_metric_2016}, with sensitivity greater than 75\% at $10^5$ false positive voxels. Derivative-SER reaches sensitivity 40\% at that level, outperforming SER as already proved.


\section{Conclusions}
This paper presents promising results for challenging NME breast lesion detection in DCE-MRI. We propose an approach that develops a linear expansion of features for every voxel in the image based on ICA, allowing for a \emph{multicurve} characterization of the enhancement behavior, in contrast with usual single-curve voxel characterization. The data-driven obtained features are used to train and test an SVM with satisfactory performance. In addition, previously the imbalanced nature of the interest class features limited automatic detection by supervised methods as SVM. In this work, we propose  parameter optimization on the SVM hyperplane location, such that the false positive rate is controlled, thus providing a solution to the low specificity problem in CAD of breast cancer. With that optimization the DSC value is increased approximately a 50\% from the default $d=0$ margin value, reaching a peak value of 0.5295, and .

\section{Acknowledgments}

This work has received funding from the European Union’s Horizon 2020 research and innovation programme under the Marie Skłodowska-Curie grant agreement No 656886,, the Austrian Nationalbank 'Jubilaeumsfond' Project Nr. 16219, the 2020 - Research and Innovation Framework Programme PHC-11-2015 Nr. 667211-2, and seed grants from Siemens Austria, Novomed, and Guerbet, France. Katja Pinker also received support from the NIH/NCI Cancer Center Support Grant P30 CA008748. We also want to thank Elena G. Avidad for her design contribution.

\section{Conflicts of interest}

The author(s) declare(s) that there is no conflict of interest regarding the publication of this paper.
\bibliographystyle{abbrv}
\bibliography{references.bib}

%

\end{document}